\documentclass[aip,jmp,graphicx]{revtex4-1}
\usepackage{graphicx}

\draft 

\begin{document}


\title[On two-body problem in spaces of constant curvature]{On the separation of variables
into relative and center of mass  motion for two-body system in
three--dimensional spaces of constant curvature.} 



\author{Yu. Kurochkin}
\email[]{y.kurochkin@dragon.bas-net.bys}
\altaffiliation[Also at ]{Belarusian National Technical University,\\
65 Nezalezhnasci av., Minsk, 220013, Belarus}
\author{Dz. Shoukavy}
\email[]{shoukavy@ifanbel.bas-net.by}
\affiliation{Institute of Physics, National Academy of Sciences of Belarus\\  68
Nezalezhnasci av., Minsk, 220072, Belarus}

\author{I. Boyarina}
\affiliation{Belarusian State Agrarian Technical University,99 Nezalezhnasci av., Minsk, 220023, Belarus }


\date{\today}

\begin{abstract}
Expressions for variables of the center of mass and relative motions
for two-body system with  different and equal masses in
three--dimensional spaces of constant curvature are introduced in
the terms  of biquaternions. The problem  of  the separation  of
center mass and relative motion variables   for action of two
particles into biquaternionic form is formulated. We showed that the
algebraic nature of these nonseparable variables follows from the
fact that algebra of biquaternions is noncommutative.  Some special
cases of separation of center mass and relative motion variables are
considered.
\end{abstract}


\maketitle 

\section{Introduction}
Quantum--mechanical problem of the particle motion in the field of
force defined by central potential on three-dimensional spherical
space $S_3$ (positive constant curvature) has been  first considered
by E. Schr\"odinger\cite{Schr} and A. Stivenson \cite{St}.  The
similar problem in the three-dimensional Lobachevsky space $^1S_3$
(negative constant curvature) has been solved by Infeld and Shild
for thr first time \cite{Inf}. Since spherical and hyperbolic spaces
have different geometrical and physical properties, most authors
investigate physical systems on spherical and hyperbolic spaces in a
separate ways. However, we believe that for certain problems a
generalized approach is more effective for both spaces.

The quantum--mechanical models based on the geometry of spaces of
constant curvature have attracted considerable attention due to
opportunity of their applications to physical problems as well as
their interesting mathematical features
(see\cite{MyR,San1,San2,San3,Bon,Grit,Sal,Vil,Otc,Red}). For
example, the model based on the Coulomb interaction on the sphere
has been used for description of excited states of excitons in
quantum dots (see \cite{Grit,Sal}). The existence of Landau levels
for the moving charged particle in  curved space were studied in
\cite{Fak,Kim}. Among other applications of non-Euclidean geometry
in theoretical physics we can highlight the usage of hyperbolic
geometry to solve  problems of relativistic kinematics \cite{Fiz}
and  the theory of relativistic nuclear collisions \cite{Bald}.
Thus,  the technique of non-Euclidean geometry is useful and
effective for a wide range of kinematic and dynamic problems.

Also, we  note that  studying of quantum mechanics on spaces of
constant curvature is important, not only for a development of our
knowledge of certain fundamental features of quantum mechanics, but
also  for convenient construction of generalized relativistic
theory.

The problem of two interacting particles  in  spaces of  constant
curvature is much more complicated because of the fact that
variables of the relative motion and center mass   in Schr\"odinger
equation can't be separated  even for a central potential of
interaction \cite{Sc1,Ot}. In the initial stage there is a problem
with the definition of the center mass in spaces of constant
curvature (see \cite{Sc1}). However, the use of vectors  of spaces
constant curvature  defined in \cite{I} provides  the opportunity to
introduce the  mentioned variables by analogy with four--velocity
(four--momentum) space  in the relativistic kinematics. Such
approach opens obvious methodical advantages and facilitates a
search for some special cases in which  complete separation of
variables is possible  that is important for practical sense.

\section{Coordinates of particles on three--dimensional sphere and  Lobachevsky space}

Three–-dimensional   spaces of constant curvature of Riemann and
Lobachevsky   can be embedded into the four--dimensional Euclidean
and pseudo-Euclidean space described by biquaternions
correspondingly as
\begin{equation}
\label{eq:1} X=iX_0+\underline{X},~ \mbox{where $i^2=\pm1$}.
\end{equation}
The case  $i^2=1$ i.e. biquaternions over double numbers corresponds
to Euclidean space, and $i^2=-1$ (biquaternions  over complex
numbers) --  pseudo--Euclidean space.

It should be noted that below in all formulas the upper sign will
refer to $S_3$ space while the lower sign to $^1S_3$ space.

The method of the unified description of the geometry  of
three–-dimensional   spaces of constant curvature belongs to
Clifford. This method is provided by choice of special form of the
biquaternions $X=-\bar{X}^*$, known as biquaternions of Minkovsky
type in  Synge's terminology in case of the pseudo-Euclidean
four--space. Here $\bar{X}=iX_0-\underline{X}$ is  a biquaternion
conjugate of a biquaternion $X$ and  the symbol $^*$  denotes
conjugate in system of double and complex numbers correspondingly.

These biquaternions obey the standard multiplicative rule
\begin{equation}
\label{eq:2}
Y=(iX'_0+\underline{X}')(iX_0+\underline{X})=\pm
X_0'X_0-(\underline{X}~\underline{X}')+iX_0'\underline{X}+
iX_0\underline{X}'+[\underline{X}'~\underline{X}],
\end{equation}
where  underline denotes  a three--dimensional vector in right part
of (\ref{eq:2}), parentheses means scalar product of
three--dimensional vectors and square brackets are cross vector of
these vectors.

The uniform equation of three--dimensional surfaces  on which
realized  Riemannian  space with constant curvature $1/R^2$ or
Lobachevsky space with constant curvature $-1/R^2$ can be written
into biquaternionic form as
\begin{equation}
\label{eq:3} X~\bar{X}=X_1^2+X_2^2+X_3^2\pm X_0^2=\pm R^2.
\end{equation}
We require the condition $X_0 >0$ for Lobachevsky space.

 The expression
(\ref{eq:3}) is invariant under transformation
\begin{equation}
\label{eq3n}
X'=AX\bar{A}^*,
\end{equation}
where $A\bar{A}=1, A^*\bar{A}^*=1$.  The set of biquaternions
(\ref{eq:1}) is invariant   under the transformation groups of
motions (\ref{eq3n})  in considered spaces. Transformations
(\ref{eq3n}) generate the   $SO(4,R)$ , $SO(3,1)$ groups
correspondingly.

Below we take  $R=1$ for  the  practical reasons.

Let us consider the motion of  two noninteracting particles in both
spaces $(S_3 $ and $^1S_3)$. The particles coordinates in embedding
four--dimensional space are represented as components of
biquaternions
\begin{equation}
\label{eq:4} X^{(1)}=iX^{(1)}_0+\underline{X}^{(1)}, \quad
X^{(2)}=iX^{(2)}_0+\underline{X}^{(2)}.
\end{equation}
Using  condition (\ref{eq:3}) we have
\begin{equation}
\label{eq:5}
X^{(1)}~\bar{X}^{(1)}=\pm 1,~~X^{(2)}~\bar{X}^{(2)}=\pm 1.
\end{equation}
These coordinates of particles are not independent.

We will use the Beltramy coordinates which are components of vectors
on sphere as independent coordinates
\begin{equation}
\label{eq:6} {\underline{q}}^{(1)}=\pm
i\frac{{\underline{X}}^{(1)}}{X_{0}^{(1)}},~{\underline{q}}^{(2)}=
\pm i\frac{{\underline{X}}^{(2)}}{X_{0}^{(2)}}.
\end{equation}

For a demonstration of the certain analogy between
three--dimensional description of the considered curvature spaces
and three--dimensional Euclidean space we give the formula for
coordinates of particles $X^{(1)}, X^{(2)}$, which are expressed in
terms of the Beltramy coordinates
\begin{equation}
\label{eq:6new}X^{(1)}=\frac{1+q^{(1)}}{\sqrt{1+\underline{q}^{(1)}\bar{\underline{q}}^{(1)}}},~
X^{(2)}=\frac{1+q^{(2)}}{\sqrt{1+\underline{q}^{(2)}\bar{\underline{q}}^{(2)}}}.
\end{equation}

We note that   definition of vectors  (\ref{eq:6}) automatically
leads to an identification of opposite points on a sphere and
thereby these  vectors belong to elliptic space. Therefore the usage
of vectors  (\ref{eq:6}) for description of the particles motion  on
the sphere  demands to take into account  this property. However we
can use biquaternions (\ref{eq:4}) to avoid these difficulties.
Nevertheless, in parallel with quaternionic variables we will use
vectors (\ref {eq:6}) since the problem for elliptic space has
independent value.

\section{Variables of the center of mass and relative
motions for  two  particle system in three--dimensional spaces of
constant curvature}

The coordinates of the center of mass for two particles with masses
$m_1$  and $m_2$ in a biquaternionic form   can be written as
\begin{equation}
\label{eq:8}
X_c=\frac{m_1X^{(1)}+m_2X^{(2)}}{\sqrt{\pm\left(m_1X^{(1)}+m_2X^{(2)}\right)\left(m_1\bar{X}^{(1)}+m_2\bar{X}^{(2)}\right)}}.
\end{equation}
We note that in this case three--dimensional coordinates
(\ref{eq:6}) of the center of mass are  components of vector
\begin{equation}
\label{eq:9} \underline{q}_c=\pm i\frac{\underline{X}_c}{X_{0c}}=\pm
i\frac{m_1\underline{X}^{(1)}+m_2\underline{X}^{(2)}}{m_1X_0^{(1)}+m_2X_0^{(2)}},
~
X_c=\frac{1+\underline{q}_c}{\sqrt{1+\underline{q}_{~c}~\bar{\underline{q}}_{~c}}}.
\end{equation}

This expression in  variables (\ref{eq:6})  has the form
\begin{equation} \label{eq:10}
\underline{q}_c=\frac{m_1\underline{q}^{(1)}/\sqrt{1+\underline{q}^{(1)}\bar{\underline{q}}^{(1)}}+m_2\underline{q}^{(2)}/\sqrt{1+\underline{q}^{(2)}\bar{\underline{q}}^{(2)}}}
{m_1/\sqrt{1+\underline{q}^{(1)}\bar{\underline{q}}^{(1)}}+m_2/\sqrt{1+\underline{q}^{(2)}\bar{\underline{q}}^{(2)}}}.
\end{equation}

It is easy to see that expression (\ref{eq:10}) for coordinates of
the center of mass coincides with correspondence expression for
three--dimensional flat space where masses  $m_1,~m_2$ replaced by
$m_1\rightarrow
m_1/\sqrt{1+\underline{q}^{(1)}\bar{\underline{q}}^{(1)}},~
m_2\rightarrow
m_2/\sqrt{1+\underline{q}^{(2)}\bar{\underline{q}}^{(2)}}$.

We note that the feature of the definition of the center mass in our
formalism in contrast with other authors (\cite{Sc1}) is manifestly
covariant under group of transformations (\ref{eq3n}).

In the paper \cite{GT} it was first shown that the model based on
the space with conformally  flat geometry  with the  similar
coordinate dependence of the mass  can be used for an explanation of
quark confinement. At the same time quantum--mechanical model based
on the geometry of the three--dimensional sphere $S_3$ has been used
for description of excitations in dimensional quantum dot  in the
work \cite{Grit} This model provides confinement of quasiparticles
and it is interpreted in terms of geometry of three-dimensional
Euclidian spaces.

Biquaternionic analog of the variables of relative motion for
two-body system is
\begin{equation}
\label{eq:11}
Y_{12}=\pm X^{(2)}~\bar{X}^{(1)},
\end{equation}
defined from
\begin{equation}
\label{eq:12}
X^{(2)}=Y_{12}~{X}^{(1)}.
\end{equation}
The independent three--dimensional coordinates of the relative
motion defined as components of relative motion vector:
\begin{eqnarray}
\label{eq:13}
q_{y}=\frac{Y_{12}-\bar{Y}_{12}}{Y_{12}+\bar{Y}_{12}}=\left<\pm
i\frac{\underline{X}^{(2)}}{X_0^{(2)}},~\mp
i\frac{\underline{X}^{(1)}}{X_0^{(1)}}\right>
=\left<\underline{q}^{(2)},~-\underline{q}^{(1)}\right>,
\end{eqnarray}
where angle brackets mean addition   rule for three--dimensional
vectors in Riemannian and Lobachevsky spaces
\begin{equation}
\label{add} {\underline{q}}''=\left<
\underline{q},\underline{q}'\right>=\frac{\underline{q}+\underline{q}'+\left[\underline{q}~\underline{q}'\right]}
{1-\left(\underline{q}~\underline{q}'\right)}.
\end{equation}
 The formula (\ref{add}) is an algebraic expression for
addition of vectors in the elliptic and Lobachevsky
three--dimensional spaces or a triangle rule (see Appendix A or more
detailed in \cite{b1,b2}). Three vectors over double numbers
correspond to ordered pairs of points (or direct lines) in the
elliptic three--space. Three vectors over complex numbers correspond
to ordered pairs of points (or direct lines) in the extended
three--dimensional Lobachevsky space. The geometry of spaces of
Riemann and Lobachevsky can be derived from properties of vectors
(\ref{eq:13}) and formula (\ref{add}).

Now let us  introduce four--dimensional  $Y_1,Y_2$ and
three--dimensional  $q_y^{(1)}, q_y^{(2)}$ coordinates with respect
to center of mass that are defined identically (\ref{eq:11}) and
(\ref{eq:12}) as
\begin{equation}
\label{eq:14}
X^{(1)}=Y_{1}~{X}_{c},~\bar{X}^{(1)}={\bar{X}}_{c}~\bar{Y}_{1}
\end{equation}
meanwhile
\begin{equation}
\label{eq:15}
{Y}_{1}=\pm X^{(1)}{\bar{X}}_{c},
\end{equation}
and  respectively
\begin{equation}
\label{eq:16}
X^{(2)}=Y_{2}~{X}_{c},~\bar{X}^{(2)}={\bar{X}}_{c}~\bar{Y}_{2},
\end{equation}
\begin{equation}
\label{eq:17}
{Y}_{2}=\pm X^{(2)}{\bar{X}}_{c}.
\end{equation}
It is clear that
\begin{equation}
\label{eq:18}
{Y}_{12}=Y_2~\bar{Y}_1.
\end{equation}
Then for the first particle we have
\begin{equation}
\label{eq:19} {\underline{q}}^{(1)}=\pm
i\frac{{\underline{X}}^{(1)}}{X_{0}^{(1)}}=\left<\frac{\underline{q}_y}{1+\frac{m_1}{m_2}\sqrt{1+\underline{q}_{y}~\underline{\bar{q}}_{y}}},
~{\underline{q}_c}\right>=\left<\underline{q}^{(1)}_{y},{\underline{q}_c}\right>,
\end{equation}
and for the second particle
\begin{equation}
\label{eq:20} {\underline{q}}^{(2)}=\pm
i\frac{{\underline{X}}^{(2)}}{X_{0}^{(2)}}=\left<\frac{\underline{q}_y}{1+\frac{m_2}{m_1}\sqrt{1+\underline{q}_y~\underline{\bar{q}}_y}},
~{\underline{q}_c}\right>=\left<\underline{q}^{(2)}_{y},{\underline{q}_c}\right>.
\end{equation}

From formula  (\ref{eq:18})  follows
\begin{equation}
\label{eq:21}
\underline{q}_y=<q_2,~-q_1>=\left<\underline{q}_y^{(2)},~-\underline{q}_y^{(1)}\right>.
\end{equation}
The introduced variables satisfy  the following conditions
\begin{equation}
\label{eq:22}
X_c~\bar{X}_c=\pm1,~Y_{12}~\bar{Y}_{12}=1,~Y_{1}~\bar{Y}_{1}=1,~Y_{2}~\bar{Y}_{2}=1.
\end{equation}

\section{Classical non--relativistic problem. The separation of variables in an action.}
An action of the  two--body problem,  which  forces of interaction
depend on the relative variable, in both spaces can be written as
\begin{equation}
\label{eq:23}
W_{12}=\int L dt= \int
\left[\frac{1}{2}\left(m_1\dot{X}^{(1)}\dot{\bar{X}}^{(1)}+m_2\dot{X}^{(2)}\dot{\bar{X}}^{(2)}\right)-V(Y_{12})
\right]dt,
\end{equation}
here L is Lagrange function and  the dot over symbols indicate the
time derivative.

The expression (\ref{eq:23}) will have a standard form  if we
substitute independent variables $\underline{q}^{(1)}$ and
$\underline{q}^{(2)}$. In this case we obtain
\begin{equation}
\label{eq:24}
W=\int\left[\frac{1}{2}\left(m_1g_{ab}(\underline{q}^{(1)})\dot{q}_a^{(1)}\dot{q}_b^{(1)}+m_2g_{ab}
(\underline{q}^{(2)})\dot{q}_a^{(2)}\dot{q}_b^{(2)}\right)-\phi(q_{12})\right]dt,
\end{equation}
where
\begin{equation}
\label{eq:25}
 g_{ab}=\frac{1}{\left(1+q^2\right)}\left(\delta_{ab}+\frac{q_aq_b}{1+q^2}\right)
\end{equation}
is a metric tensor of a sphere written in the variables which are
components of vectors on sphere. However,  feature of this work is
that we use action (\ref{eq:23}) written  in terms of
four--dimensional biquaternionic variables taking into account the
additional conditions (\ref{eq:5}) and (\ref{eq:22}).

Taking account  expressions (\ref{eq:14}), (\ref{eq:16}) for
$X^{(1)}$, $X^{(2)}$ and  condition (\ref{eq:22}) the expression
(\ref{eq:23}) can be transformed into
\begin{eqnarray}
\label{eq:26}
\label{d} W=\frac{1}{2}\int \left[m_1\left(\pm
\dot{Y}_1\dot{\bar{Y}}_1\pm\dot{X}_c \dot{\bar{X}}_c\right)
+m_2\left(\pm\dot{Y}_2\dot{\bar{Y}}_2\pm\dot{X}_c
\dot{\bar{X}}_c\right)+m_1\left({Y}_1
\dot{X}_c\bar{X}_c\dot{\bar{Y}}_1+\dot{Y}_1{X}_c\dot{\bar{X}}_c\bar{Y}_1
\right)\right.\nonumber
\\
\left.+m_2\left({Y}_2
\dot{X}_c\bar{X}_c\dot{\bar{Y}}_2+\dot{Y}_2{X}_c\dot{\bar{X}}_c\bar{Y}_2
\right)+2V(Y_{12})\right]dt,\qquad\qquad
\end{eqnarray}

Variables of the center mass and relative motions in the first two
terms of expression of a kinetic part (\ref{eq:26}) are separated.
The following terms have  variables both relative motion and center
of mass. The separation of variables in these terms is impossible
because of noncommutative biquaternions.  Thus, the nature of
inseparable variables of  the center mass and relative motions can
be explained by non-commutative biquaternions in algebraic sense.

\section{Some special cases}
Nevertheless, formula (\ref{eq:26}) is very convenient for searching
of some special cases where complete separation of variables of
relative motion and the motion of the center of mass is possible.
For example, it is clear that we can separate variables when
$\dot{{X}}_{c}=0$. In this case two massive points move along
two-dimensional spheres  and  lie on the same straight line
(geodesic line) round  the general center of mass. This case known
as the dumbbell--shaped figure \cite{Sc1}. The distance from general
center to points defined by masses ratio.

Let us rewrite a kinetic part of subintegral function (\ref{eq:26})
using operator $Y_{12}$ (\ref{eq:11}) as follows

\begin{eqnarray}
\label{eq:27}
L'=\pm\frac{m_1m_2\dot{Y}_{12}\dot{\bar{Y}}_{12}(m_1+m_2)}{m_1^2+m_2^2+2m_1m_2Y_{12(0)}}\mp
\frac{m_1^2m_2^2\dot{Y}^2_{12(0)}(m_1+m_2)}{\left(m_1^2+m_2^2+2m_1m_2Y_{12(0)}\right)^2}
\pm (m_1+m_2)\dot{X}_c\dot{\bar{X}}_c\qquad\qquad\nonumber\\
+\frac{m_1m_2}{m_1^2+m_2^2+2m_1m_2Y_{12(0)}}\left[m_1\left(Y_{12}\dot{X}_c
\bar{X}_c \dot{\bar{Y}}_{12}+\dot{Y}_{12}X_c \dot{\bar{X}}_c
\bar{Y}_{12}\right)+m_2\left(\bar{Y}_{12}\dot{X}_c
\bar{X}_c \dot{Y}_{12}+\dot{\bar{Y}}_{12}X_c \dot{\bar{X}}_c
Y_{12}\right)\right] \nonumber\\
+\frac{m_1m_2}{m_1^2+m_2^2+2m_1m_2Y_{12(0)}}\left[m_1\left(\dot{X}_c
\bar{X}_c \dot{Y}_{12}+\dot{\bar{Y}}_{12}X_c
\dot{\bar{X}}_c\right)+m_2\left(\dot{X}_c\bar{X}_c\dot{\bar{Y}}_{12}+\dot{Y_{12}}X_c\dot{\bar{X}}_c\right)\right]. \qquad
\end{eqnarray}

Setting $Y_{12}=\cos{r}+\underline{n}\sin{r}$, where $r$ is a
distances between points and $n^2=n^{2*}=1$ for Riemannian space we
find
\begin{eqnarray}
\label{eq:28}
L'=\frac{m_1m_2(m_1+m_2)}{F}\Big(\dot{r}^2+\underline{\dot{n}}^2\sin^2{r}+(\dot{\underline{n}}\dot{X}_c\bar{X}_c\underline{n}-\underline{n}\dot{X}_c\bar{X}_c\dot{\underline{n}})\sin^2{r}\Big)
\qquad\qquad\nonumber\\
-\frac{m_1^2m_2^2(m_1+m_2)\sin^2{r}\dot{r}^2}{F^2}+(m_1+m_2)\dot{X}_c\dot{\bar{X}}_c+\frac{m_1m_2(m_1-m_2)}{F}\Big[(\underline{n}\dot{X}_c\bar{X}_c+\dot{X}_c\bar{X}_c\underline{n})\dot{r}
\nonumber\\
(\underline{\dot{n}}\dot{X}_c\bar{X}_c+\dot{X}_c\bar{X}_c\underline{\dot{n}})\sin{r}\cos{r}-(\underline{n}\dot{X}_c\bar{X}_c+\dot{X}_c\bar{X}_c\underline{n})\dot{r}\cos{r}
-(\underline{\dot{n}}\dot{X}_c\bar{X}_c+\dot{X}_c\bar{X}_c\underline{\dot{n}})\sin{r}\Big], \qquad
\end{eqnarray}
where $F=m_1^2+m_2^2+2m_1m_2\cos{r}$.

In the case of equal masses $m_1=m_2=m$ we get
\begin{eqnarray}
\label{eq:29}
L'={m}\left[2\sin^2{\frac{r}{2}}\left(\underline{\dot{n}}^2+\dot{\underline{n}}\dot{X}_c\bar{X}_c\underline{n}-\underline{n}\dot{X}_c\bar{X}_c\dot{\underline{n}}\right)
+\frac{\dot{r}^2}{2}+2\dot{X}_c\dot{\bar{X}}_c \right].
\end{eqnarray}

By analogy with the Riemannian space we use the substitution
$Y_{12}= \cosh{r}+\underline{n}\sinh{r}$, where $n^2=n^{2*}=-1$ for
Lobachevsky space as result we get the kinetic part of subintegral
function $L'$ as
\begin{eqnarray}
\label{eq:30}
L'=\frac{m_1m_2(m_1+m_2)}{F_L}\Big(\dot{r}^2-\underline{\dot{n}}^2\sinh^2{r}+(\dot{\underline{n}}\dot{X}_c\bar{X}_c\underline{n}-\underline{n}\dot{X}_c\bar{X}_c\dot{\underline{n}})\sinh^2{r}\Big)
-(m_1+m_2)\dot{X}_c\dot{\bar{X}}_c\nonumber\\
+\frac{m_1^2m_2^2(m_1+m_2)\sinh^2{r}\dot{r}^2}{F_L^2}+\frac{m_1m_2(m_1-m_2)}{F}\Big[(\underline{\dot{n}}\dot{X}_c\bar{X}_c+\dot{X}_c\bar{X}_c\underline{\dot{n}})\sinh{r}\cosh{r}
\qquad\nonumber\\
+(\underline{n}\dot{X}_c\bar{X}_c+\dot{X}_c\bar{X}_c\underline{n})\dot{r}-(\underline{n}\dot{X}_c\bar{X}_c+\dot{X}_c\bar{X}_c\underline{n})\dot{r}\cosh{r}-(\underline{\dot{n}}\dot{X}_c\bar{X}_c
+\dot{X}_c\bar{X}_c\underline{\dot{n}})\sinh{r}\Big],\qquad
\end{eqnarray}
where $F_L=m_1^2+m_2^2+2m_1m_2\cosh{r}$.

The formula (\ref{eq:30}) may be rewritten for equal masses
$m_1=m_2=m$ as follows
\begin{eqnarray}
\label{eq:31}
L'={m}\left[2\sinh^2{\frac{r}{2}}\left(-\underline{\dot{n}}^2+\dot{\underline{n}}\dot{X}_c\bar{X}_c\underline{n}-\underline{n}\dot{X}_c\bar{X}_c\dot{\underline{n}}\right)
+\frac{\dot{r}^2}{2}-2\dot{X}_c\dot{\bar{X}}_c \right].
\end{eqnarray}

We note that taking into account  the homogeneity of
three--dimensional spaces of constant curvature $S_3$ and $^1S_3$ we
can choose the center of mass as a reference point. Then we get
\begin{eqnarray}
\label{eq:32} X_c=(i, \underline{0}),~
\dot{X}_c=(0,\underline{\dot{X}}),\qquad\qquad\qquad\nonumber\\
\underline{n}=i\underline{\hat{n}}~ \mbox{where}~ (i=\pm1),~
\underline{\hat{n}}^{*}=\underline{\hat{n}},~\underline{\hat{n}}~\underline{\bar{\hat{n}}}=\underline{\hat{n}}^2=1,
\end{eqnarray}
and structure of the expressions (\ref{eq:27})--(\ref{eq:31}) is
more understandable.

We  also should remark that as far as  astrophysical data don't
exclude the existence of a small curvature of the spatial part of
the Universe, the expressions (28)--(31) might be taken into account
under calculations of the motion of interacting  Galaxies pair.

Let us consider  the case when relative distance is much smaller
than the radius of curvature, i.e. $r\ll1$. In this case we can
neglect  terms that contain $r^2$ and so, hence  $\cos r\approx
1,~\sin{r}\approx r$ for Riemannian space and $\cosh r\approx
1,~\sinh{r}\approx r$ for Lobachevsky geometry. As a  result, the
kinetic part of subintegral function for both spaces takes the form
\begin{equation}
\label{eq:33}
L'=\frac{m_1m_2}{m_1+m_2}\dot{r}^2\pm(m_1+m_2)\dot{X}_c\dot{\bar{X}}_c.
\end{equation}

Thus, separation of variables of the center of mass and the relative
motion for two-body system in spaces of constant curvature is
possible  when relative distance is much smaller than the radius of
curvature i.e. we ignore the  rotational degree of freedom.

Nevertheless, expression (\ref{eq:33}) is the very rough
approximation to formulas (\ref{eq:28}), (\ref{eq:30}). It
corresponds to the symmetry of the s-states in quantum mechanical
Coulomb problems on three--dimensional sphere and Lobachevsky space
and some configurations which are investigated in cosmology.

\section{Conclusion}
The  problem  of  the separation  of center of mass  and relative
motion variables  is formulated  for the case of two particles in
the terms of biquaternions. We showed that an algebraic nature of
these nonseparable variables follows from the fact that  algebra of
biquaternions is  noncommutative. It is shown that complete
separation of variables of the relative motions and center of mass
for two-body problem in spaces of constant curvature is possible
only for  zero approximation. In this approximation rotation degrees
of the freedom of the system are excluded.

\appendix

\section{Appendixes}
Let us three arbitrary points (vertexes of the triangle) on the
three dimensional sphere (or hyperboloid) defined by coordinates
which are the components of biquaternios $X^{(1)}, X^{(2)}, X^{(3)}$
correspondingly. We take the all notations of the articles.

We introduce  biquaternions as  transformations $X^{(1)}\rightarrow
X^{(2)}, X^{(2)}\rightarrow X^{(3)}, X^{(3)}\rightarrow X^{(1)}$ as
follows $Q_{12}=X^{(2)}\bar{X}^{(1)}, Q_{23}=X^{(3)}\bar{X}^{(2)},
Q_{31}=X^{(1)}\bar{X}^{(3)}$.

We have $Q_{12}\bar{Q}_{12} =1, Q_{23}\bar{Q}_{23} =1,
Q_{31}\bar{Q}_{31} =1$ in corresponds with formulas (\ref{eq:3}) and
(\ref{eq:5}). Let us write $Q_{12}, Q_{23}$ and $Q_{31}$  in the
form which automatic satisfy by these conditions. That is
$$
Q_{12}=\frac{1+\underline{q}_{12}}{\sqrt{1+\underline{q}_{12}\underline{\bar{q}}_{12}}},
Q_{23}=\frac{1+\underline{q}_{23}}{\sqrt{1+\underline{q}_{23}\underline{\bar{q}}_{23}}},
Q_{31}=\frac{1+\underline{q}_{31}}{\sqrt{1+\underline{q}_{31}\underline{\bar{q}}_{31}}}.
$$

You can easily check that $Q_{31}=Q_{23}Q_{12}$ then using the
standard multiplicative rule (\ref{eq:2}) we obtain

$$
Q_{31}=\frac{1+\underline{q}_{31}}{\sqrt{1+\underline{q}_{31}\underline{\bar{q}}_{31}}}=
\frac{1-\left(\underline{q}_{12}\underline{q}_{23}\right)+\underline{q}_{12}+\underline{q}_{23}+\left[\underline{q}_{12}\underline{q}_{23}\right]}
{\sqrt{1+\underline{q}_{12}\underline{\bar{q}}_{12}}~\sqrt{1+\underline{q}_{23}\underline{\bar{q}}_{23}}}.
$$
Thus, for the scalar part we have

$$
\frac{1}{\sqrt{1+\underline{q}_{31}\underline{\bar{q}}_{31}}}=\frac{1-\left(\underline{q}_{12}\underline{q}_{23}\right)}{\sqrt{1+\underline{q}_{12}\underline{\bar{q}}_{12}}
~\sqrt{1+\underline{q}_{23}\underline{\bar{q}}_{23}}},
$$
and for vector part we get
$$
\frac{\underline{q}_{31}}{\sqrt{1+\underline{q}_{31}\underline{\bar{q}}_{31}}}=\frac{\underline{q}_{12}+\underline{q}_{23}+\left[\underline{q}_{12}\underline{q}_{23}\right]}{\sqrt{1+\underline{q}_{12}\underline{\bar{q}}_{12}}
~\sqrt{1+\underline{q}_{23}\underline{\bar{q}}_{23}}}.
$$

Divide the vector part by scalar one to get the formula (\ref{add})
of the article.

The vectors in formula (7) corresponds to pairs points, or
transformations $X^0 \rightarrow X^1, X^0 \rightarrow X^2, X^0
\rightarrow X^3$, where $X^0=i$. Thus $Q_{01}=X^{(1)}\bar{X}^{(0)}$,
$Q_{02}=X^{(2)}\bar{X}^{(0)}$, $Q_{03}=X^{(3)}\bar{X}^{(0)}$ and

$$Q_{01}=\frac{1+q_{01}}{\sqrt{1+\underline{q}_{01}\underline{\bar{q}}_{01}}}, Q_{02}=\frac{1+q_{02}}{\sqrt{1+\underline{q}_{02}\underline{\bar{q}}_{02}}},
Q_{03}=\frac{1+q_{03}}{\sqrt{1+\underline{q}_{03}\underline{\bar{q}}_{03}}},
\mbox{where}$$
 $$q_{0s}=\pm i \frac{\underline{X}^{(s)}}{X_0^{(s)}}, s=1,2,3.$$

Because $Q_{12}=Q_{02}\overline{Q}_{01}=\pm X^{(2)}\bar{X}^{(1)},
Q_{23}=Q_{03}\overline{Q}_{02}=\pm X^{(3)}\bar{X}^{(2)},
Q_{31}=Q_{01}\overline{Q}_{03}=\pm X^{(1)}\bar{X}^{(3)}$ we have
formula  (\ref{eq:13}) by analog with above derivation for
transformations of the arbitrary points.

%

\end{document}